# Selective data collection in vehicular networks for traffic control applications

Bartłomiej Płaczek

*Faculty of Transport, Silesian University of Technology*
*Krasińskiego 8, 40-019 Katowice, Poland*
*Tel.: +48 0326034243 Fax: +48 0326034288*
*E-mail address: bartlomiej.placzek@polsl.pl*

**Abstract**

Vehicular sensor network (VSN) is an emerging technology, which combines wireless communication offered by vehicular ad hoc networks (VANET) with sensing devices installed in vehicles. VSN creates a huge opportunity to extend the road-side sensor infrastructure of existing traffic control systems. The efficient use of the wireless communication medium is one of the basic issues in VSN applications development. This paper introduces a novel method of selective data collection for traffic control applications, which provides a significant reduction in data amounts transmitted through VSN. The underlying idea is to detect the necessity of data transfers on the basis of uncertainty determination of the traffic control decisions. According to the proposed approach, sensor data are transmitted from vehicles to the control node only at selected time moments. Data collected in VSN are processed using on-line traffic simulation technique, which enables traffic flow prediction, performance evaluation of control strategies and uncertainty estimation. If precision of the resulting information is insufficient, the optimal control strategy cannot be derived without ambiguity. As a result the control decision becomes uncertain and it is a signal informing that new traffic data from VSN are necessary to provide more precise prediction and to reduce the uncertainty of decision. The proposed method can be applied in traffic control systems of different types e.g. traffic signals, variable speed limits, and dynamic route guidance. The effectiveness of this method is illustrated in an experimental study on traffic control at signalised intersection.

## 1. Introduction

Vehicular ad hoc networks (VANETs) provide wireless communication between vehicles and their environment for dynamic traffic data transfers with a low cost and high accuracy. Numerous VANETs applications are assumed to contribute to a safer, more efficient and more comfortable driving experience (Toor et al. 2008). Vehicular sensor network (VSN) is an emerging technology, which combines the aforementioned wireless communication offered by VANET with sensing devices installed in vehicles (Lee and Gerla, 2010). Sensors available in vehicles can gather data sets including localisations, speeds, directions, accelerations, etc. The vehicles participating in VSN can be used as the sources of information to determine accurately the traffic flow characteristics.

The road traffic control becomes an important application area of VSNs. This new technology creates a huge opportunity to extend the road-side sensor infrastructure of the existing traffic control and monitoring systems (Lee and Gerla, 2010). A major drawback of the current-generation systems is a limited access to the traffic data that are usually registered only in few fixed positions (Płaczek and Staniek, 2007). The coverage of these sensing platforms is narrow due to high installation and maintenance costs. It is expected that the VSNs will help to overcome these limitations as a remarkable feature of the sensor network is the capability to monitor every single vehicle dynamically.

For many cases the scope of real-time vehicular data available in VSN exceeds the requirements of particular traffic control implementations. Moreover, the transfer of all data records from vehicles to the traffic control unit is highly not advisable due to the bandwidth-limited wireless communication medium. Number of vehicles in VSN may use the same transmission medium for many applications of different purposes (e.g. control, safety, comfort). In dense road traffic the periodic transmissions may consume the entire channel bandwidth resulting in excessive congestion and delays in the communication network. Therefore the efficient use of the wireless communication channel is one of the basic issues in VSNs applications development (Lee and Gerla, 2010).

The contribution of this paper is the proposal of a novel method of selective data collection for traffic control applications, which provides a significant reduction in data amounts transmitted through VSN. The underlying idea is to detect the necessity of data transfers on the basis of uncertainty determination of the traffic control decisions. Fig. 1 presents the overall design of the road traffic control application in VSN environment. This schema includes the data collection module, which sends queries to retrieve useful traffic data from the network. The advantage of the introduced approach is that it uses on-time queries (Sun, 2007) instead of periodical data sampling. On-time queries are generated only at the specific moments of time, when a high uncertainty level of the control decision occurs. A task of the decision making module is to use the currently available information in a selection of optimal traffic control strategy among several alternatives (e.g. optimal selection of travel route or traffic signal timing).

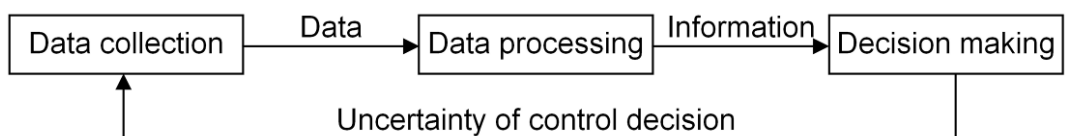

Fig. 1. Functional modules of traffic control application

The uncertainty level of control decisions depends on the precision of the information that is achieved using the data processing unit. The data processing procedure involves short term traffic predictions to evaluate the performance of alternative control strategies that can be selected for implementation at a given moment. If the precision of the resulting information is insufficient, the optimal control strategy cannot be derived without ambiguity. As a result the control decision becomes uncertain and it is a signal for the data collection unit informing that new traffic data from VSN are necessary in the system to provide more precise prediction and to reduce the uncertainty of decision.

The rest of the paper is organised as follows: Related works are reviewed and analysed in Section 2. Section 3 describes the basic steps of decision making process in the traffic control application and discuss the impact of data collection on the decision uncertainty. A predictive traffic model for on-line data processing is presented in Section 4. In Section 5, the algorithm of selective data collection is introduced in details. Section 6 contains the results of an experimental study on data collection in VSN for the traffic control at a signalised intersection. Finally, in Section 7, conclusions are given and some future research directions are outlined.

## 2. Related works

The traffic monitoring is one of the most important services in VSNs which brings benefits to road traffic control applications. Many research and implementation efforts have been involved in the traffic data collection (e.g., Amin et al. 2008; Hull et al. 2006; Lo et al. 2008). The traffic data collection platforms are typically based on client-server architectures. Such platforms usually consist of servers and vehicles equipped with GPS modules and the wireless communication interfaces, such as 3G or WiFi networks. The sensed data (e.g. the speed and the position) are sent to the server for traffic monitoring. Clearly, with these data from VSN, traffic state can be estimated without any aid of costly road-side sensors for traffic monitoring systems. Using the communication capability of each vehicle, the traffic control applications are supplied with up-to-date information about particular events that effect control decisions.

The emergence of VSNs technologies has made it possible to use novel, more effective techniques to deal with the problems of road traffic control. Several traffic control algorithms have been developed in this field of research for signalised intersections. Most methods are based on wireless communication between vehicles and road-side control nodes (e.g., Abishek et al. 2009; Gradinescu et al. 2007; Wenjie et al. 2005). These adaptive signal control schemes use real-time sensor data collected from vehicles (e.g. their positions and speeds) to minimise travel time and delay experienced by drivers at road intersections.

Numerous works have been devoted to the problem of traffic control in road networks. In (Collins and Muntean 2008) a traffic management system has been introduced that includes a server-side decision making module for optimal route selection in urban network and enables the dissemination of instructions to vehicles. Wedde et al. (2007) have proposed a routing scheme from multi-agent routing algorithm to control road traffic. In that approach vehicles are directed under decentralised control at each road intersection. The technique proposed by Inoue et al. (2007) employs traffic information sharing and route selection procedures to address the problem of vehicle traffic congestion. On the basis of shared traffic information, congestion free routes are selected. In the Street Smart Scheme (Dornbush and Joshi, 2007), each vehicle builds a speed map based on the speed of other vehicles in its vicinity and transmits it to the neighbouring vehicles. As a result, each vehicle is able to select the fastest route. In (Mohandas et al. 2009) an approach has been proposed to deal with the problem of traffic congestion using a congestion control algorithm designed for the Internet. A strategy developed by Wang et al. (2007) attends the problem of traffic congestion in intersections where a ramp leads on to a highway. According to this proactive traffic merging strategy, vehicles share their velocity and acceleration information with other vehicles in the network. Each vehicle decides where and when it can merge onto the highway before arriving at the merging point. Another traffic control application enabled in VSN is dynamic speed control including speed warning, variable speed limits, and cooperative driving (Chuah et al. 2008).

Although much work has been done to develop VSNs applications for road traffic control, little research has examined their requirements on input data. In most of the above cited studies, the real-time sensed data are assumed to be delivered continuously from all vehicles in a certain area to the control node. Such periodical data sampling scheme may cause excessive congestion and latency in the communication network due to the bandwidth-limited wireless communication medium. Therefore, more research is needed to determine required input data sets as well as sampling rates that are necessary for the decision making in traffic control applications. On-line evaluation of these characteristics will enable reduction in data amounts transmitted through VSN.

The problem of data congestion and transfer latency in VANETs has been addressed in many publications (e.g., Cuckov and Song, 2010; Hossain et al. 2010; Hung and Peng 2010; Saleet and Basir, 2007; Salhi et al. 2009). A way to avoid this problem is to increase the bandwidth but usually this is not cost-effective because quite often the congestion arises due to the inappropriate allocation or utilisation of resources (Hossain et al. 2010). Another approach is to fairly share the available bandwidth among the users without causing any congestion, i.e. congestion avoidance with efficient bandwidth management. Most of researches tend to find improvements using effective medium access control protocols (Sikdar, 2008), transmission strategies (e.g. data dissemination (Cuckov and Song, 2010), data aggregation (Saleet and Basir, 2007), self-organisation mechanisms (Salhi et al. 2009)), and wireless technologies (Lee and Gerla, 2010) in order to reduce the congestion and latency. While above solutions focus on the expansion of the available network capacity, the selective data collection method introduced in this paper contributes to reduction of data congestion in VSN by minimising the demand for data transmission.

In the literature several methods have been introduced for wireless sensor networks that enable the optimisation of data collection procedures. Spatial and temporal suppression based techniques have been demonstrated to be useful in reducing the amount of sensor data transmitted for monitoring physical phenomena (Kulik et al. 2008; Puggioni and Gelfand, 2010; Silberstein et al. 2006). The underlying insight is that different observed states of the physical phenomena are in fact temporally as well as spatially correlated. Temporal suppression is the most basic method: sensor readings are transmitted only from those nodes where a change occurred since the last transmission (Reis et al. 2009). Spatial suppression includes methods such as clustered aggregation (Min and Chung, 2010) and model-based suppression (Chu et al. 2006). They aim to reduce redundant transmissions by exploiting the spatial correlation of sensor readings. If the sensor readings of neighbouring sensor nodes are the same or similar, the transmission of those sensed values can be suppressed. In (Silberstein et al. 2006) a combined spatio-temporal suppression algorithm was introduced that considers the node readings and their differences along transmission paths to suppress reports from individual nodes.

The suppression based methods use a subset of sensor readings from selected nodes to derive actual values of the monitored parameters for all remaining nodes in the network. However, in those methods the data collection procedure is executed at regular time intervals because the sensor readings have to be continuously analysed to suppress the unnecessary transmissions from particular nodes. The method introduced in this paper enables the time selective execution of the data collection procedure in VSN. According to the proposed approach, sensor data are transmitted from nodes (i.e. vehicles) to the control node only at selected time moments. For the remaining time periods, the data collection procedure is ceased and the control node approximates all the sensor readings on its own, using a predictive traffic model. It should be noted here that in such time periods the control node does not communicate with other nodes (vehicles). This approach exploits the fact that traffic control applications can tolerate approximate sensor readings. Nevertheless, the uncertainty of traffic parameters approximation (prediction) has to be appropriately low to ensure the optimal performance of a traffic control application. Thus, the uncertainty of this approximation is taken into account to decide when the data collection has to be executed.

## 3. Uncertainty estimation of traffic control decisions

This section discusses basic steps of the decision making process in a traffic control application. A formal description of a generalised traffic control procedure is introduced using set theory and relational representation of uncertainty. The impact of data collection on the decision uncertainty is analysed in this context. On this basis a principle is formulated that enables the selection of time moments for data collection operations.

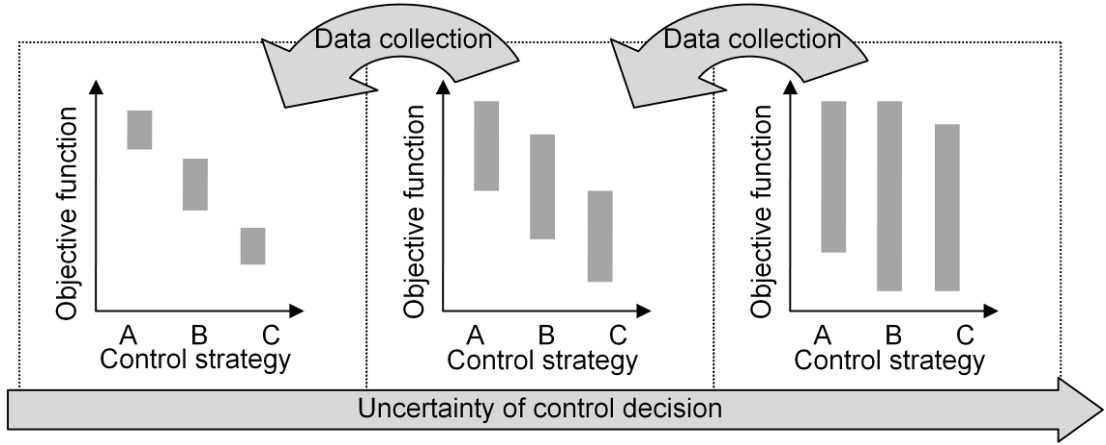

Fig. 2. Impact of data collection on decision uncertainty

The objective function of traffic control optimisation problem can be any combination of the following performance measures: average delay per vehicle, maximum individual delay, percentage of cars that are stopped, average number of stops, throughput of intersections, travel time, etc. Prediction of objective function values for all applicable strategies is required to decide the optimal control strategy. This prediction is always uncertain due to the nondeterministic nature of traffic flow. The additional source of uncertainty in the traffic control is an estimation of the current traffic state. These two types of uncertainty affect control decisions. However, the estimation uncertainty can be reduced by delivering new data collected in VSN. An example of the results of data collection operations and their impact on the control decision uncertainty is presented in fig. 2. The question to be answered here is how to recognise the situations in which the additional data from VSN are necessary to make optimal control decisions.

To address the above issue let us consider the road traffic control procedure as an iterative process that is based on control decisions making in successive time intervals. Each control decision determines the selection of control strategy, which will be implemented in a given moment of time. A set of consecutive decisions creates a traffic control programme for the defined time period $T$. The objective of traffic control procedure is to find an optimal programme, which minimises the objective function.

We will assume that the control decisions are made at some time steps, thus the period $T$ can be defined as a set: $T = \{t_0, t_0+1, ..., t_n\}$. The control decision for time moment $t$ will be denoted by $d(t)$. It can be interpreted as a mapping d: $T \rightarrow W$, where $W = \{w_i\}$ is a set of all applicable control strategies. A relation $D_t \subseteq T \times W$ will be used to describe the traffic control programme. At a given time moment $t \in T$ the $d(t_i)$ are already made decisions for $t_i \leq t$, and for $t_i > t$ they are decisions to be made in the future. Thus, the control programme at time step $t$ has the following properties:

$$\forall t_i \leq t \left( (t_i, w) \in D_t \Rightarrow w = d(t_i) \right), \ \forall t_i > t \ \forall w \in W \ (t_i, w) \in D_t \quad (1)$$

Traffic data will be represented using a family of relations $(F_t)_{t \in T}$, $F_t \subseteq T \times R$, where $R$ is a multidimensional, discrete space of traffic flow parameters. The family of relations has to be used here because the traffic data may be updated at each time step. Some portions of these relations are determined on the basis of traffic data collected in VSN, but only for selected time moments in the past. The remaining part of traffic data is a result of a traffic prediction.

Values of the objective function are predicted in the proposed approach using a model of traffic flow. A knowledge which is included in the model can be described using relation $E \subseteq (R \times W)^{(n+1)} \times \mathbf{R}$, where $\mathbf{R}$ denotes the set of real numbers. At a time moment $t \in T$ values of the objective function can be predicted on the basis of data aggregated in relations $F_t$ and $E$, using the following formula:

$$\hat{E}_t = \{x \in \mathbf{R} : (r_0, w_0, r_0, w_0, \ldots, r_n, w_n, x) \in E \land \forall i \in \{0,1,\ldots,n\}((t_i.r_i) \in F_t) \land (t_i, w_i) \in D_t\} . \quad (2)$$

By $\hat{E}_t^{d(t)}$ we will denote the values set of objective function predicted at a time moment $t \in T$ for the case when the control decision $d(t)$ is made. It means that $(t, w) \in D_t \Rightarrow w = d(t)$. We will say that the decision $d'(t)$ is more effective than decision $d(t)$ if the following inequality of probabilities holds:

$$P(e' < e) > P(e' > e), \quad (3)$$

where: $e' \in \hat{E}_t^{d'(t)}$, $e \in \hat{E}_t^{d(t)}$, and the probabilities are computed accordingly:

$$P(e' < e) = \frac{|\{(e',e) \in \hat{E}_t^{d'(t)} \times \hat{E}_t^{d(t)} : e' < e\}|}{|\hat{E}_t^{d'(t)}| \cdot |\hat{E}_t^{d(t)}|},$$

$$P(e' > e) = \frac{|\{(e',e) \in \hat{E}_t^{d'(t)} \times \hat{E}_t^{d(t)} : e' > e\}|}{|\hat{E}_t^{d'(t)}| \cdot |\hat{E}_t^{d(t)}|}. \quad (4)$$

To simplify notation, let $L[d'(t), d(t)] = d'(t)$ denote that the condition (3) is satisfied i.e. decision $d'(t)$ is more effective than $d(t)$. Uncertainty of this conclusion will be determined using the following formula:

$$UNC_{L[d'(t),d(t)]} = 1 - P(e' < e) + P(e' > e). \quad (5)$$

Above conclusion is certain if $P(e' < e) = 1$. Uncertainty is zero in that case, because $P(e' < e) + P(e' > e) \leq 1$ is always true. The uncertainty value increases when the probability $P(e' < e)$ falls, however this value is always lower than one since the condition (3) has to be satisfied.

A control decision $d^*(t)$ is an optimal one at a time moment $t$ if

$$\forall d(t) \in W \left( d(t) \neq d^*(t) \Rightarrow L[d^*(t), d(t)] = d^*(t) \right), \quad (6)$$

and uncertainty associated with this decision is given by:

$$UNC_{d^*(t)} = \max_{d(t) \in W - d^*(t)} \{UNC_{L[d^*(t),d(t)]}\}. \quad (7)$$

The control decision is based on evaluation of the objective function (2) for all applicable strategies. If the objective function can be evaluated precisely, then the uncertainty level of control decision is low. It means that the uncertainty depends directly on cardinality of set $\hat{E}_t$. Using the equation (2) it can be found that the cardinality of set $\hat{E}_t$ is proportional to the cardinality of relation $F_t$. Thus, in most cases the uncertainty of control decision can be reduced by decreasing the cardinality of relation $F_t$. This reduction is achieved through updating the traffic data in $F_t$. The updating operation involves traffic data collection and prediction. The data collection is necessary to determine current traffic parameters. Moreover, the collected data are further used as a starting point of the traffic prediction and allow the prediction to be made with more confidence.

The main motivation for introducing the selective data collection in VSN is the observation that different monitored parameters of the road traffic are temporally as well as spatially correlated. Hence, the traffic data collected in VSN do not change abruptly and randomly between two time steps of the traffic control procedure but instead changes occur in predictable patterns. The proposed method is based on an assumption that data collection frequency can be optimised using a predictive traffic model. However, the uncertainty of traffic prediction has to be appropriately low to ensure the optimal performance of traffic control procedure. Thus, the uncertainty monitoring is necessary to adjust the data collection frequency accordingly.

### 4. Predictive traffic model for on-line data processing

The traffic control procedure defined in previous section requires an application of a traffic model to aggregate and process the data collected in VSN. The traffic model is necessary for several tasks, including traffic flow prediction, performance evaluation of control strategies and uncertainty estimation. In the proposed method the fuzzy cellular model is applied to on-line simulation of traffic flow. The on-line simulation technique enables rapid evaluation of alternate courses of action in order to aid in decision making processes (Kosonen, 2003). It allows us to determine the current traffic parameters (real time simulation) as well as to predict the future state of the traffic flow (faster than real time simulation). The term on-line means that the simulation is synchronised with real time and it is adjusted to traffic data collected in VSN.

The fuzzy cellular model of road traffic was intended for on-line simulation and satisfies specific requirements of traffic control applications (Płaczek, 2010). This model is based on cellular automata approach to traffic modelling that ensures accurate simulation of the real traffic phenomena (Nagel and Schreckenberg, 1992). To deal with nondeterministic traffic processes the uncertainty is described in the cellular model using fuzzy sets theory. All parameters of vehicles are represented individually by fuzzy numbers. These facts along with low computational complexity make the model suitable for on-line processing of traffic data.

A traffic lane in the fuzzy cellular model is divided into cells that correspond to the road segments of equal length. The traffic state is described in discrete time steps. These two basic assumptions are consistent with those of the Nagel-Schreckenberg cellular automata model. Thus, the calibration methods proposed by Nagel and Schreckenberg (1992) are also applicable here for determination of the cells length and vehicles properties. A novel feature distinguishing this approach from the other cellular models is that vehicle position, its velocity and other parameters are modelled by fuzzy numbers defined on the set of integers. Moreover, also the rule of model transition from one time step to the next is based on fuzzy definitions of basic arithmetical operations.

In order to reduce the computational effort associated with on-line simulation, the fuzzy cellular model was implemented using the concept of ordered fuzzy numbers (Kosiński, 2006). The algebra of ordered fuzzy numbers is a significantly more efficient tool than the solution based on classical fuzzy numbers and extension principle applied in (Płaczek, 2010). Hereinafter, all the ordered fuzzy numbers are represented by four integers and the following notation is used: $A = (a_1, a_2, a_3, a_4)$. The necessary arithmetical operations, i.e. difference, sum and maximum of the ordered fuzzy numbers, are computed as follows:

$$A + B = (a_1, a_2, a_3, a_4) + (b_1, b_2, b_3, b_4) = (a_1 + b_1, a_2 + b_2, a_3 + b_3, a_4 + b_4),$$
$$A - B = (a_1, a_2, a_3, a_4) - (b_1, b_2, b_3, b_4) = (a_1 - b_1, a_2 - b_2, a_3 - b_3, a_4 - b_4), \quad (8)$$
$$\min\{A, B\} = \min\{(a_1, a_2, a_3, a_4), (b_1, b_2, b_3, b_4)\}$$
$$= (\min\{a_1, b_1\}, \min\{a_2, b_2\}, \min\{a_3, b_3\}, \min\{a_4, b_4\})$$

Road traffic stream is represented in the fuzzy cellular model as a set of vehicles. A vehicle $n$ is described by its position $X_{n,t}$, velocity $V_{n,t}$ (in cells per time step), maximal velocity $V_n^{max}$ and acceleration $A_n$. All these quantities are expressed by fuzzy numbers. The position $X_{n,t}$ is a fuzzy number defined on the set of cells indexes. Velocity of vehicle $n$ at time step $t$ is computed as follows:

$$V_{n,t} = \min\{V_{n,t-1} + A_n(V_{n,t-1}), G_{n,t}, V_n^{\max}\}. \qquad (9)$$

Acceleration is defined as a function of velocity to enable implementation of a slow-to-stop rule that exhibits more realistic microscopic driver behaviour (Clarridge and Salomaa, 2009). $G_{n,t}$ is the fuzzy number of free cells in front of a vehicle $n$:

$$G_{n,t} = X_{n-1,t} - X_{n,t} - (1,1,1,1), \qquad (10)$$

where $n - 1$ denotes the number corresponding to the lead vehicle and $n$ that of the following vehicle. If there is no lead vehicle in front of the vehicle $n$ then $G_{n,t}$ is assumed to be equal to $V_n^{max}$.

After determination of velocities for all vehicles, their positions are updated. The position of the vehicle $n$ at the next time step $(t+1)$ is computed on the basis of the model state at time $t$:

$$X_{n,t+1} = X_{n,t} + V_{n,t}, \qquad (11)$$

The preceding formulation of the fuzzy cellular model is illustrated in fig. 3, which shows the results of numerical motion simulation of two accelerating vehicles for five time steps. At the first time step of the simulation vehicles are stopped in the first and second cell. The maximal velocity of vehicles in this example is set as follows: $V_1^{\max} = V_2^{\max} = (1, 2, 2, 3)$. The acceleration for both vehicles can take two fuzzy values depending on the vehicle's velocity:

$$A_n(V_{n,t-1}) = \begin{cases} (1,1,1,1), & V_{n,t-1} = V_n^{\max} \text{ or } V_{n,t-1} = V_n^{\max} - (1,0,0,0) \\ (0,1,1,1), & \text{else} \end{cases}. \qquad (12)$$

The fuzzy cellular model is implemented for predicting the performance of control strategies (objective function). Thus in the presented simulation environment it is necessary to provide algorithms for computations of the performance measures. To this end a function $S_C$ is defined that acts on directed fuzzy numbers:

$$S_C(A) = (s_1, s_2, s_3, s_4), s_i = \begin{cases} 0, & |A_c| < 5 - i \\ 1, & |A_c| \geq 5 - i \end{cases}, \qquad (13)$$

where $A_C$ is a set of integers used in notation of the fuzzy number $A$, which satisfy condition denoted by $C$:

$$A_C = \{a_i \in \{a_1, a_2, a_3, a_4\} : a_i \text{ satisfies condition } C\}, \qquad (14)$$

Function $S_C(A)$ allows us to determine a level of confidence that the condition $C$ is satisfied by $A$. The value of this function is (0, 0, 0, 0) if the condition $C$ is false for $A$ and (1, 1, 1, 1) if the condition is true. Any other combination of $s_i$ means that the condition is partially satisfied by $A$. For example, let us check the condition "vehicle is stopped" for the fourth time step of the simulation presented in fig. 3. Such a condition can be written as $V_{n,4} = 0$ and the corresponding form of the function (13) is $S_{=0}(V_{n,4})$. The values of this function are as follows: $S_{=0}(V_{1,4}) = (0, 0, 0, 0)$ for the lead vehicle and $S_{=0}(V_{2,4}) = (0, 0, 0, 1)$ for the following vehicle. It means that the lead vehicle is not stopped and there is a possibility that following vehicle is stopped.

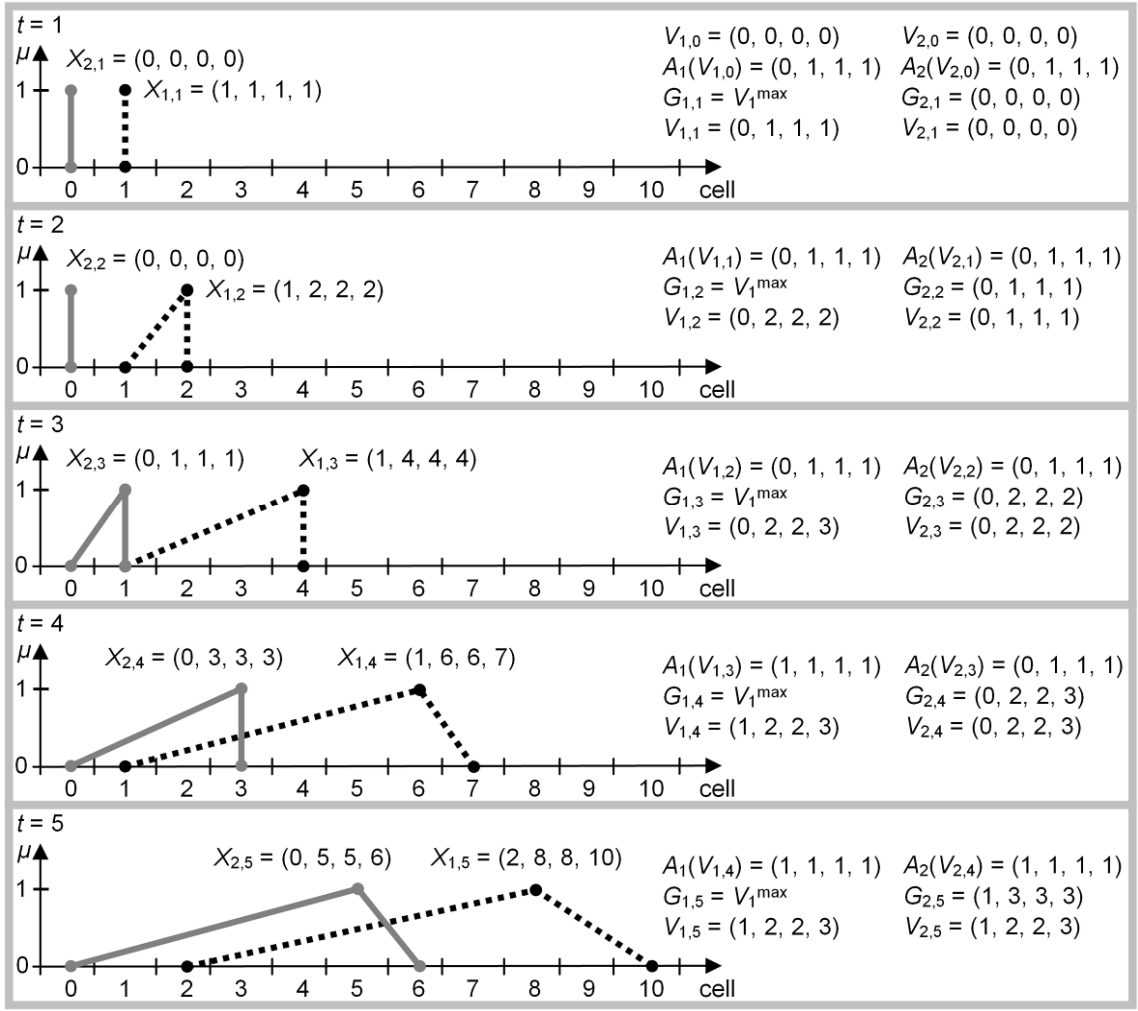

Fig. 3. Motion simulation of two vehicles

Traffic performance measures can be formulated in terms of the fuzzy cellular model, using the function $S_C$. The most commonly used criteria of performance are average delay and average number of stops. Average delay in time steps per vehicle is defined as:

$$\overline{delay} = \tfrac{1}{N} \sum_n \sum_t S_{=0}(V_{n,t}), \qquad (15)$$

where $N$ is the total number of vehicles. And average number of stops is given by:

$$\overline{stops} = \tfrac{1}{N} \sum_n \sum_t \min\{S_{>0}(V_{n,t-1}), S_{=0}(V_{n,t})\}. \qquad (16)$$

The quotient $A/N$ of an ordered fuzzy number $A$ and an integer $N$ is computed according to formula: $A/N = (a_1/N, a_2/N, a_3/N, a_4/N)$. Additionally, the values of $a_i/N$ are rounded to the nearest integers.

The main advantage of the predictive model presented in this section relies on the fact that the traffic prediction is computationally efficient and the uncertainty of the results is taken into account. The results concerning traffic performance are represented by means of fuzzy numbers. As it is shown in the next section, this representation is convenient for the determination of uncertainty in control decisions.

## 5. Uncertainty dependent data collection in traffic control application

The data collection algorithm is discussed in this section as a component of a traffic control procedure. Scope of the real-time data that have to be collected in VSN depends directly on the traffic model used for on-line simulation. In case of the fuzzy cellular model considered above it is required to provide data on the location of vehicles. Optional data that can be used to fine tune parameters of this traffic model include velocity and class of vehicles.

The operation of data collection is initialised by traffic control node, which sends query to the VSN data management system. Such system mimics classical database management systems. Thus, VSN can be viewed as a relation of tuples that are distributed across the nodes (vehicles) (Chu et al. 2006). The VSN data management system processes the query and collects necessary data readings reported by vehicles in the network. Finally, the result of query is sent back to the traffic control node.

The proposed algorithm (fig.4) provides time selective data collection in the sense that queries are generated at selected time steps (iterations) of the traffic control procedure. An uncertainty threshold is used to decide if new data have to be collected at a given time step. Namely, the data collection is executed only when the level of uncertainty, evaluated using equation (7), exceeds a predetermined uncertainty threshold. This technique enables the considerable reduction of number of queries that are necessary to capture sufficient information for the traffic control optimisation.

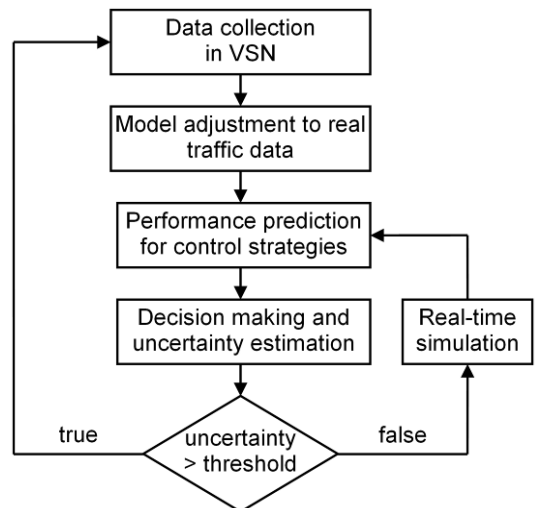

Fig. 4. Overview of uncertainty dependent data collection algorithm

After each new traffic data delivery from VSN the traffic model is appropriately adjusted to mirror the current traffic situation as accurately as possible. To facilitate on-line simulation, the received traffic data must be processed to maintain consistency between simulated and measured traffic. The model adjustment operation includes the generation and removal of vehicles, adjusting their position, and fine tuning maximal velocity and acceleration parameters. The location data has to be translated into terms of cells that are basic units used to describe vehicles positions in the model. The modification of the maximal velocity and acceleration is based on statistics which aggregate the results of multiple data collections. Since the on-line simulation is implemented to evaluate the performance of traffic control strategies, the model adjustment operation has to take into account also the real-time status data of traffic control operations.

The traffic performance for all applicable control strategies is predicted by faster than real-time simulation using the estimation of current traffic state to determine initial conditions.

The current traffic state is estimated directly on the basis of actual data collected in VSN only at selected time steps. In remaining cases this estimation involves the real-time simulation. The result of performance prediction obtained for each control strategy is an ordered fuzzy number representing a set of possible values of objective function (see eq. 2). The examples of performance prediction results for three alternative control strategies are presented in fig. 5.

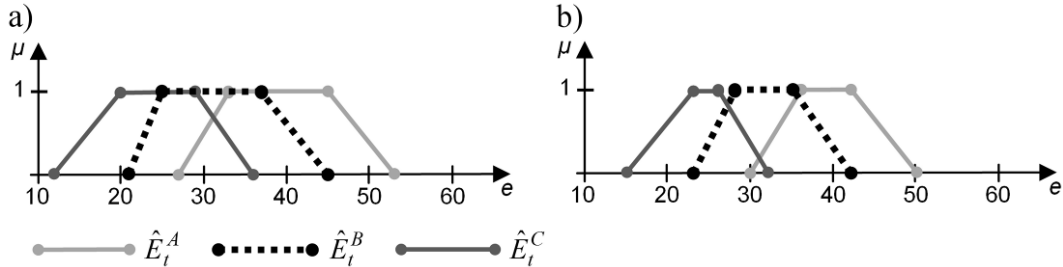

Fig. 5. Examples of performance prediction results

The optimal control strategy is selected among all alternatives during the decision making phase. The control decision is based on the evaluation of objective function (2) for all applicable strategies. The objective function can be defined as a combination of different performance measures, however for the examples presented here (fig. 5) a single measure is used, namely the average delay in seconds per vehicle. The basic issue of decision making is to select the control strategy which minimises the objective function. Thus, there is the necessity of comparing the ordered fuzzy numbers. For the proposed algorithm the probabilistic approach to fuzzy numbers comparison (Sevastianov, 2007) is employed. This method estimates quantitatively the probability to which one fuzzy number is less or equal to another fuzzy number.

The examples in fig. 5 take into account three alternative control strategies denoted by *A*, *B*, and *C*. For the first example (fig. 5 a) the corresponding predictions of the performance are represented by following fuzzy numbers:

$$\hat{E}_t^A = (12, 20, 29, 35), \quad \hat{E}_t^B = (21, 25, 38, 45), \quad \hat{E}_t^C = (27, 33, 45, 53). \quad (17)$$

Using the method outlined in (Sevastianov, 2007) the probabilities can be computed as follows: $P(e_C < e_A) = 0{,}99$ and $P(e_C > e_A) = 0{,}00$, where $P(e_X < e_Y)$ is the probability at which value of the objective function predicted for control strategy *X* is less than the value for strategy *Y*. Since the condition (3) is satisfied, the control strategy *C* is decided to be more effective than strategy *A*. This conclusion can be written in equation form as $L(C, A) = C$ and its uncertainty is determined by: $UNC_{L(C,A)} = 1 - P(e_C < e_A) + P(e_C > e_A) = 0{,}01$. The strategies *C* and *B* have to be compared in the same way: $P(e_C < e_B) = 0{,}91$, $P(e_C > e_B) = 0{,}06$, thus $L(C, B) = C$ and $UNC_{L(C,B)} = 0{,}15$. In this example the result of decision making indicates that $d^*(t) = C$ is the optimal control strategy. Using equation (7) the uncertainty of this control decision can be computed: $UNC_C = \max\{UNC_{L(C,A)}, UNC_{L(C,B)}\} = 0{,}15$.

Let us further assume that the uncertainty threshold is exceeded for the first example discussed above, and in consequence the data collection operation is activated at the next time step. The new traffic data enable more precise prediction of the control strategies performance, as shown in fig. 5 b:

$$\hat{E}_{t+1}^A = (15, 23, 26, 32), \quad \hat{E}_{t+1}^B = (24, 28, 35, 42), \quad \hat{E}_{t+1}^C = (30, 36, 42, 50). \quad (18)$$

For this prediction the new values of probabilities are computed: $P(e_C < e_A) = 1{,}00$, $P(e_C > e_A) = 0{,}00$, $P(e_C < e_B) = 0{,}98$, and $P(e_C > e_B) = 0{,}01$. Therefore, the uncertainty of control decision $d^*(t+1) = C$ is reduced in this example: $UNC_C = 0{,}03$.

The above example shows a typical scenario of uncertainty dependent data collection in the context of general definition of traffic control procedure. The proposed algorithm can be applied in traffic control systems of different types (traffic signals, variable speed limits, dynamic route guidance etc.). In the next section an experimental study is presented on signal control at an isolated intersection.

## 6. Case application and experimental results

The proposed data collection algorithm was applied to a VSN-based traffic control at a signalised intersection. The topology of intersection as well as the signal phases are illustrated in fig. 6. The traffic control procedure in this study is consistent with the general definition presented in Section 3. The procedure is executed in time steps of one second. At each time step a control decision is made regarding the selection of signal phase. The objective function used in this study is the average delay per vehicle. Thus, at each time step the signal phase is selected which minimises the delay predicted by the fuzzy cellular model.

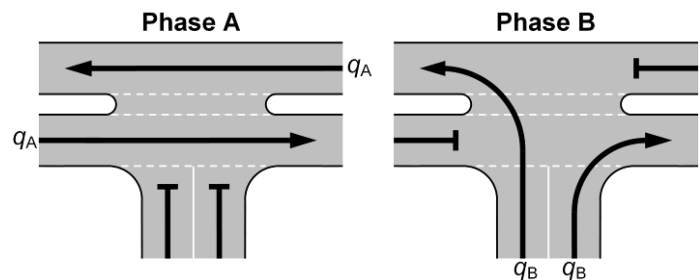

Fig. 6. Intersection topology and signal phases

The system architecture enabling the traffic control application consists of two types of VSN nodes: control node and vehicle nodes. The control node installed at the road intersection collects sensor data from the vehicle nodes and executes the traffic control procedure. Each vehicle in the system is equipped with a wireless communication unit and uses a GPS device to determine its position and speed. Every time a vehicle approaches the intersection, it has to register itself by sending a hello message to the control node. The data collection operation is initialised by the control node which generates queries to retrieve data concerning locations and velocities of all registered vehicles.

Two scenarios of the data collection are analysed in this study. The first scenario assumes that the continuous data collection is employed. It means that queries are generated at each time step of the traffic control procedure, when the signal phase selection has to be made. In the second scenario the selective data collection is applied according to the algorithm presented in the previous section. The analysis is focused on the reduction of number of queries as well as on performance of the traffic control application in the examined scenarios.

The effectiveness of the proposed solution was evaluated by using a simulation environment and realistic mobility models. Experiments were performed in VISSIM road traffic simulator. The VISSIM software provides an application programming interface (API), which was used in this study to develop an emulator of the traffic control node along with the data collection unit.

A sample of the simulation results for a time period of 1350 seconds is presented in fig. 7. The uncertainty threshold was set to 5% in this case and the traffic flow volumes at the intersection were assumed as follows: $q_A = 150$ vph and $q_B = 300$ vph. The diagram in fig. 7 shows variations of the decision uncertainty for the selective data collection scenario as well as the numbers of queries generated in the both analysed scenarios. In the first scenario a query has to be generated whenever there is a possibility to change the current signal phase (excluding the inter-green and minimum green time intervals). In the second scenario queries are generated only if the current uncertainty level of control decision exceeds the predetermined threshold. It can be seen that the number of generated queries is substantially lower for the second scenario. An important observation needs to be made at this point regarding the introduced algorithm. Fig. 7. shows that the frequency of querying in the selective data collection process is highly unstable as it is dynamically adapted to the current state of the traffic flow. Thus, the uncertainty dependent data collection cannot be simply replaced by any periodical sampling method.

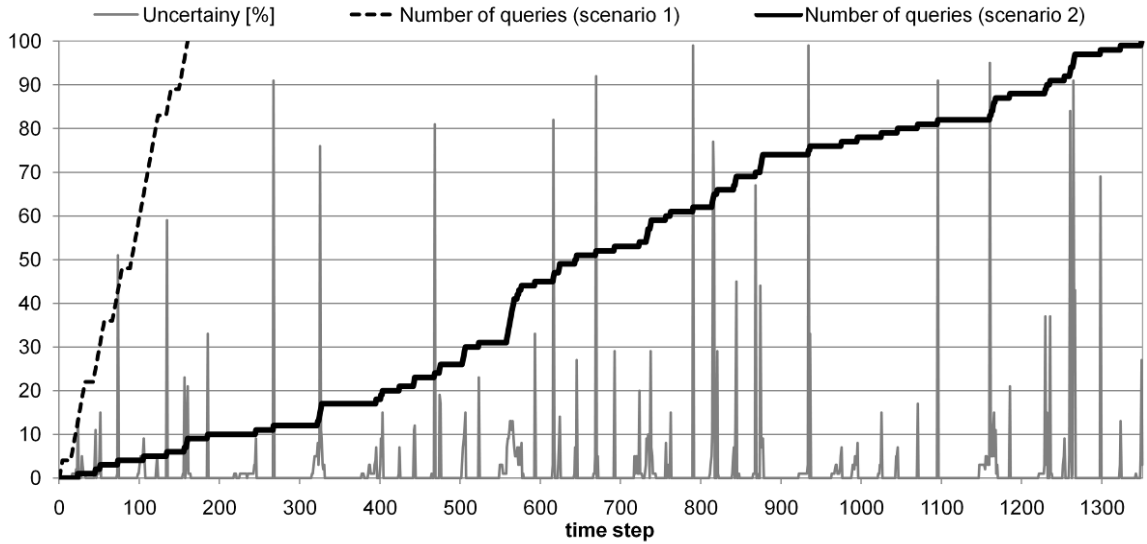

Fig. 7. Sample of simulation results

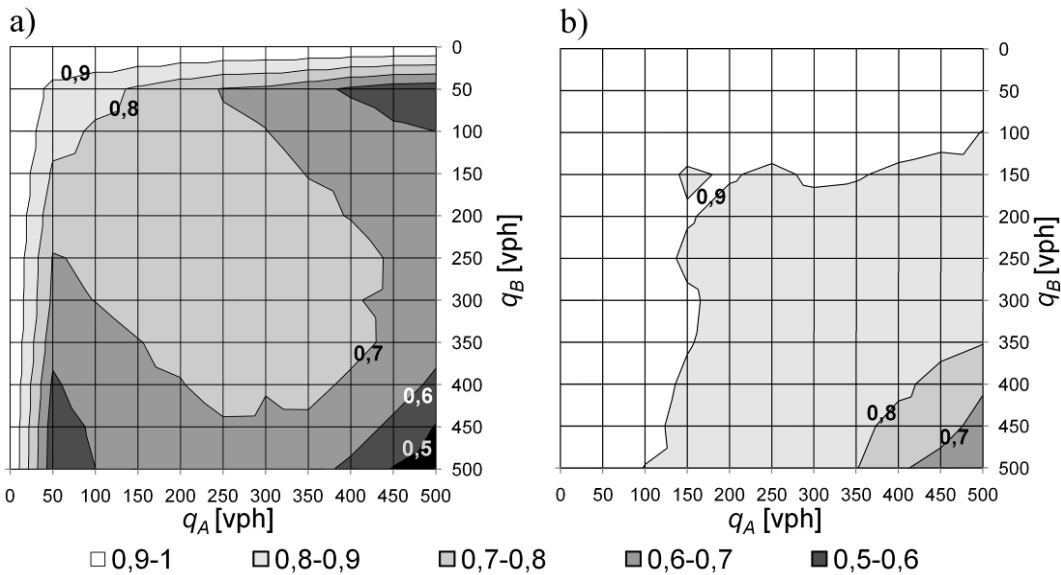

Fig. 8. Query reduction factor (*QRF*) for two values of uncertainty threshold: a) 0%, b) 10%

Contour plots in fig. 8 demonstrate the reduction in number of queries, which was observed during simulations. The query reduction factor (*QRF*) is defined as the ratio of the number of reduced queries in the selective data collection to the total number of queries for the continuous data collection: $QRF = (QN_1 - QN_2) / QN_1$, where $QN_s$ denotes the number of queries for scenario *s*. Values of the *QRF* were determined for different combinations of the traffic flow volumes. The average *QRF* was 0,74 for the uncertainty threshold 0% (fig. 8 a) and 0,90 for threshold 10% (fig. 8 b). These results confirm that the obtained reduction in number of queries is significant for a wide range of traffic conditions.

Further tests were performed to determine the effect of selective data collection on performance of the traffic control. The delay increase factor (*DIF*) was defined for this purpose as follows: $DIF = (delay_2 - delay_1) / delay_1$, where $delay_s$ is an average delay observed for scenario *s*. The results of *DIF* evaluation are presented in fig. 9. The average value of *DIF* was 0,03 for the uncertainty threshold 0% (fig. 9 a) and 0,14 for threshold 10% (fig. 9 b). The results show that the increase of delay is insubstantial in most of the analysed cases. A decline in performance is perceptible only for the high volumes of the traffic flow (congested traffic state) and it is intensified by increasing the uncertainty threshold.

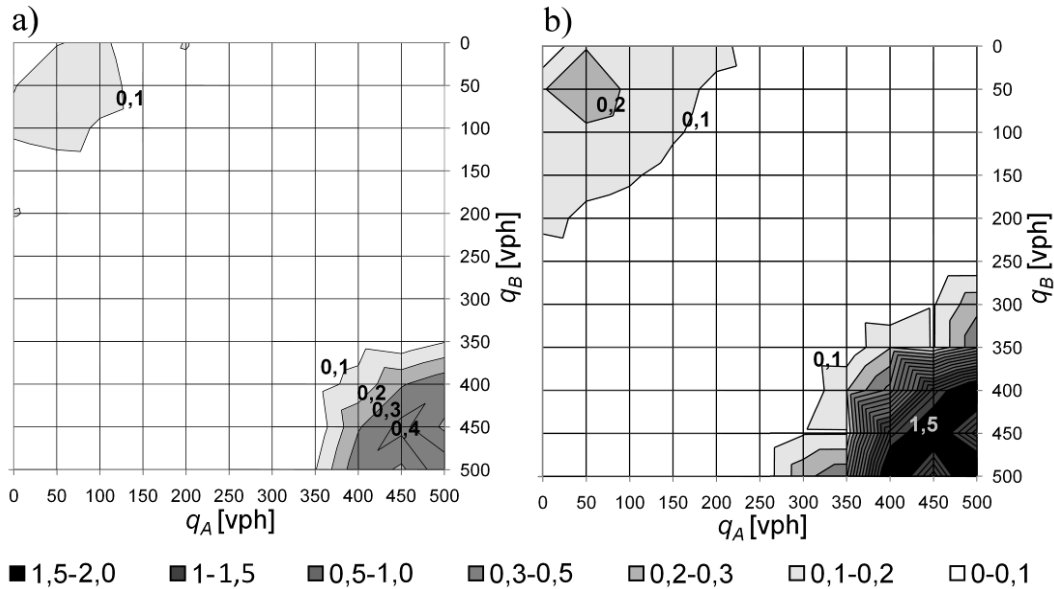

Fig. 9. Delay increase factor (*DIF*) for two values of uncertainty threshold: a) 0%, b) 10%

In fig. 10 and 11 the number of queries as well as the performance of traffic control is presented in relation to the sum of flow volumes that are used to describe the traffic approaching the intersection. Fig. 10 shows the average number of queries generated in the second scenario during a one hour period of data collection. To refer these values to the results of continuous querying, it should be noted here that the number of queries for the first scenario varies in a range between 2200 and 3600.

Fig. 11 depicts the average increase of delay, which was calculated as a difference between delays observed in the second and in the first scenario. The diagram in fig. 11 demonstrates that the uncertainty threshold values of 10% and higher cause a significant increase in the delay if the sum of traffic flow volumes exceeds 750 vph. This fact indicates that the uncertainty threshold has to be carefully determined for congested traffic conditions. Comparing the results in fig. 10 with those in fig. 11, it can be seen that there is a trade-off between the reduction of number of queries and the optimisation of traffic control performance. The trade-off can be obtained by suitably adjusting the uncertainty threshold.

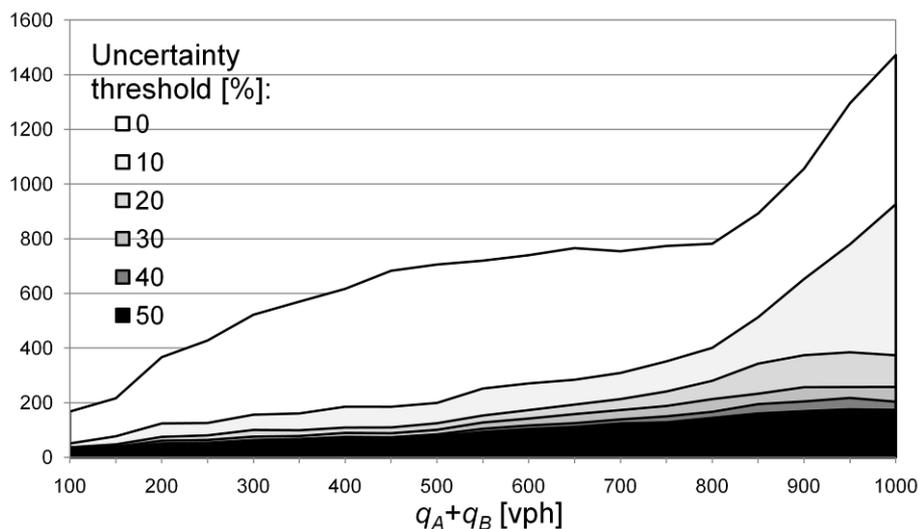

Fig. 10. Average number of queries generated over an one hour period

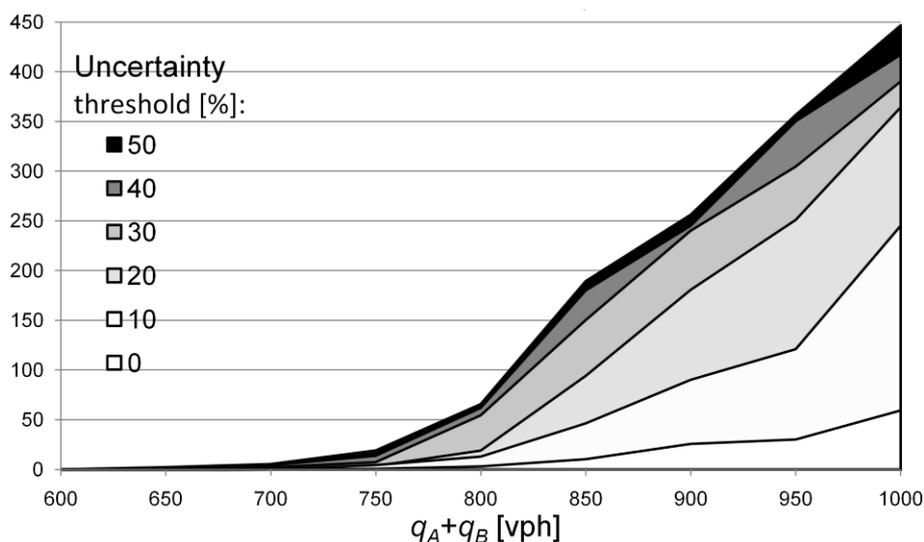

Fig. 11. Average increase of delay in seconds per vehicle

## 7. Conclusions

In this paper a novel approach was proposed to effectively reduce the amount of data that have to be transmitted through VSN in order to collect the sufficient information for traffic control applications. The offered approach is based on an original concept of the uncertainty dependent data collection. According to this concept the necessity of data transfers can be detected by the evaluation of uncertainty of traffic control decisions. The basic formulas for the uncertainty calculations were derived using a formal description of generalised traffic control procedure. The uncertainty evaluation was carried out by means of fuzzy arithmetic. In this solution, the ordered fuzzy numbers were used to represent imprecision of the available traffic information.

The proposed approach was applied to develop a new data collection algorithm for the traffic control applications in VSN environment. In this algorithm, the sensor data are transmitted from vehicles to the control node only at selected time moments. For the remaining periods of time the sensor data are replaced by the results of an on-line traffic simulation. In a typical scenario, after some number of the simulation steps the precision of resulting information decreases, the control decisions become uncertain and in consequence the new traffic data have to be collected from vehicles. These operations result in an effective data collection, which is dynamically adapted to the current state of the traffic flow. The implementation of an on-line simulation technique along with a predictive traffic model enables the significant reduction of the demand for data transmission.

The Effectiveness of the introduced data collection method was demonstrated in an experimental study on the traffic control at a signalised intersection. Experiments were carried out using a simulation environment and realistic mobility models. The experimental results confirmed that the proposed algorithm significantly reduces the number of data transmissions for a wide range of traffic flow volumes. Extensive tests were performed to determine the effect of selective data collection on performance of the traffic control. The obtained results show that the decline in performance is insubstantial in most of the analysed cases.

The approach introduced in this paper provides the foundation for further research on uncertainty dependent data collection in VANETs. An interesting issue is the possibility to integrate the proposed method with the suppression based techniques, which is expected to enable the selection of vehicle nodes that are necessary to participate in the data collection procedure. In future studies the feasibility will be explored of using the proposed method for the usefulness evaluation of particular data readings that can be delivered by the vehicle nodes.